# Quantum Hyperdense Coding for Distributed Communications

Sándor Imre

Mobile Communications and Quantum Technologies Laboratory

Budapest. Univ. of Technology

imre@hit.bme.hu

**Abstract**

Superdense coding proved that entanglement-assisted quantum communications can improve the data transmission rates compared to classical systems. It allows sending 2 classical bits between the parties in exchange of 1 quantum bit and a pre-shared entangled Bell pair. This paper introduces a new protocol which is intended for distributed communication. Using a pre-shared entangled Bell pair and 1 classical bit 2,5 classical bits can be transmitted in average. This means not only valuable increase in capacity but the two-way distributed operation opens new fields of investigation.

1. INTRODUCTION

The continuously growing capacity needs require significant effort to find new solutions for communications networks [5] at various layers of the systems. Quantum mechanics can be utilized both in communications and computing [3,7]. It is known that appropriate encoding of classical information into quantum states may improve the classical capacity of the channels. One of the earliest examples was the so called superdense coding [4]. However, the transmission rates depend not only on the capacity of the physical medium but accessing methods has great influence to, especially in distributed systems.

Various medium access control protocols have been developed [6] which can either be distributed or centralized. In the former case users have individual strategies how to access the medium. The lack of coordination results typically in low performance. The throughput of the system can be improved if the users are cooperative i.e. they adjust their strategies to achieve common goals such as fairness, maximised joint throughput, etc. Practically it requires following the same rules in each terminal. Finally, centralised MAC can provide more efficient use of the common medium, involving sophisticated operation. In this paper we consider the well-known and widely studied slotted-Aloha protocol for the sake of easier explanation of new possibilities offered by quantum mechanics based communications.

This paper is organised as follows. Having explained some motivations to apply quantum communication techniques in classical ad hoc systems in Section 1 we summarize the main properties of the classical slotted-Aloha and superdense coding used as references for the new protocol in Section 2. Section 3 introduces the proposed combined classical-quantum protocol called hyperdense coding and discusses its efficiency measured by the average throughput per timeslot. Comparisons to reference systems are also provided. Finally Section 4 concludes the paper and summarizes further research directions.

## 2. Reference protocols

The hyperdense coding protocol inherits some properties of superdense coding and slotted-Aloha. In this section both ancestors are briefly summarised to provide the theoretical background.

### 2.1 Superdense coding

By means of the postulates of quantum mechanics [3] any quantum communication protocol or algorithm can be analyzed, however it is worth emphasizing one of their consequences called *entanglement*. Entangled states such as [8]

$$|\beta_{00}\rangle = \frac{|00\rangle + |11\rangle}{\sqrt{2}} \qquad (2.1)$$

include special connection between the qubits building them. Independently from the distances of the qubits and time interval between the two measurements, measuring one of the qubits will influence the other qubit, too. Entanglement is one of the most efficient and dangerous effect in quantum computing and communications enabling e.g. teleportation, communication over zero-capacity channels and tremendously fast algorithms (thus often mentioned as a basic resource). On the other hand, entanglement with the environment stands behind the fact that quantum computers have not been left the laboratories, yet.

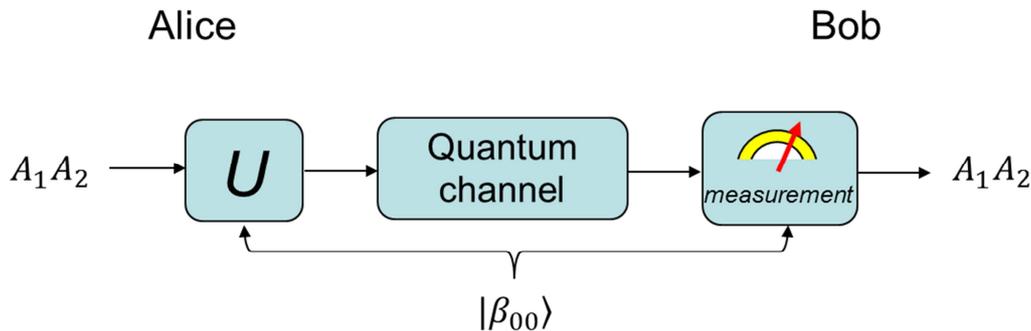

Figure 2.1 Superdense coding

In case of superdense coding [4] originally Alice sends classical bitpairs $A_1$ and $A_2$ to Bob. She encodes the information into the half of the previously shared $|\beta_{00}\rangle$ state [8] by means of four different transformations in encoder *U* (*I* and the three Pauli transforms *X,Y* and *Z*) selected by the bitpair values (see Table 2.1). Next the transformed qubit is sent to Bob who performs a measurement on the joined pair (received and locally available half-pairs). Since the four possible joint states $|\beta_{kl}\rangle$ are orthogonal Bell-states they can be trivially distinguished by means of a projective measurement $P_{kl} = |\beta_{kl}\rangle\langle\beta_{kl}|, \; k,l \in \{0,1\}$. In order to make the identification easier *A* and *B* in the formulas refer to which qubit originates from Alice and which from Bob.

| dibit | Alice's transform $U$ | joint state at Bob after he received Alice's qubit |
|---|---|---|
| 00 | $I$ | $\|\beta_{00}\rangle = \dfrac{\|00\rangle^{AB}+\|11\rangle^{AB}}{\sqrt{2}}$ |
| 01 | $Z$ | $\|\beta_{01}\rangle = \dfrac{\|00\rangle^{AB}-\|11\rangle^{AB}}{\sqrt{2}}$ |
| 10 | $X$ | $\|\beta_{10}\rangle = \dfrac{\|10\rangle^{AB}+\|01\rangle^{AB}}{\sqrt{2}}$ |
| 11 | $iY$ | $\|\beta_{11}\rangle = \dfrac{\|01\rangle^{AB}-\|10\rangle^{AB}}{\sqrt{2}}$ |

Table 2.1 Operation table of superdense coding

By means of superdense coding theoretically the classical capacity of a quantum channel can be doubled since Alice transmits 2 bits using 1 qubit.

2.2 The Classical Slotted-Aloha Protocol

The slotted-Aloha protocol is a suitable reference for medium access control protocols in distributed ad hoc networks where no coordination is available among the users in the form of a base station or access point.

In slotted-Aloha systems [1,2] the time axis consists of equal length periods for packet transmission and there is perfect synchronisation among the users, see Fig. 2.2. Each user $i$ sends its packet exactly at the beginning of the slot to the common channel with probability $p_i$ and leaves the medium empty with $1-p_i$. Assuming idealistic medium (no errors in the Physical layer) $M$ players sending packets independently from one another and using the same strategy i.e. they are cooperative, the normalized number of successfully transmitted packets per slot can be represented by a Bernoulli random variable $s_i$ with success probability

$$q_i = p_i(1-p_i)^{M-1}, \tag{2.2}$$

and expected value

$$\mathrm{E}(s_i) = q_i. \tag{2.3}$$

We are interested in the maximisation of the throughput

$$\frac{d\mathrm{E}(s_i)}{dp_i} = 0 \Rightarrow p_i = \frac{1}{M} \Rightarrow \max_{p_i} \mathrm{E}(s_i) = \frac{1}{M}\left(1-\frac{1}{M}\right)^{M-1}, \tag{2.4}$$

and for all the users $s = \underset{i=1}{\overset{M}{*}} s_i$ we get

$$\max_{p_i} E(s) = M \cdot \frac{1}{M}\left(1-\frac{1}{M}\right)^{M-1} = \left(1-\frac{1}{M}\right)^{M-1}, \qquad (2.5)$$

which is a very important MAC related result. To maximize the throughput users have to be familiar with the total number of players. The well-known exponential backoff in 802.11 is doing nothing else then trying to estimate this unknown value continuously.

Finally we show that going with $M$ to infinity

$$\lim_{M\to\infty} \max_{p_i} E(s) = \lim_{M\to\infty}\left(1-\frac{1}{M}\right)^{M-1} = \lim_{M\to\infty} \frac{1}{1-\frac{1}{M}}\left(1-\frac{1}{M}\right)^M = \frac{1}{e}. \qquad (2.6)$$

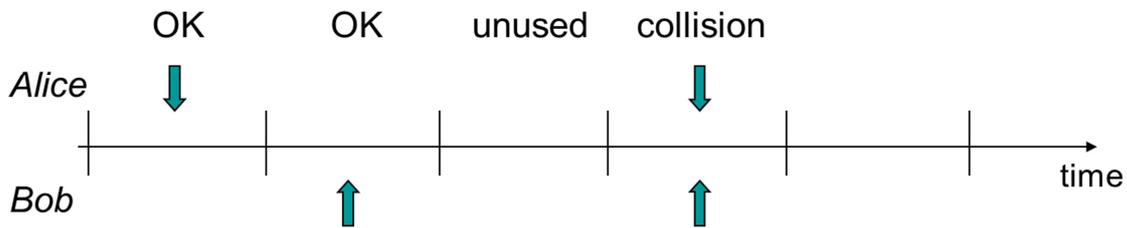

Figure 2.2 Slotted-Aloha MAC

Although we did not emphasize the above description approaches the problem from game theory point of view i.e. players optimized their individual throughput following the same strategy. Result $p_i = \frac{1}{M}$ means the Pareto optimum of the game since none of the users can increase its payoff without hurting at least one player.

Without giving the corresponding proof we would reach the same result if we summarized independent random variables $s_i$ for $M$ users by means of convolution building a binomial random variable and maximized the overall throughput and calculated the portion of one user.

3. Hyperdense coding

First we discuss the combined classical-quantum MAC protocol from Alice's point of view according to Fig. 3.1. Bob shall follow the same steps in a parallel manner.

It is assumed that Alice and Bob have already shared a $|\beta_{00}\rangle$ Bell pair before timeslot $T$ begins. For the sake of simplicity each timeslot has 1 bit duration. In a given timeslot Alice tries to transmit two classical bits $A_1$ and $A_2$.

1. Alice measures her half pair in the $\{|0\rangle, |1\rangle\}$ basis obtaining $C_A$ containing one of the two basis states with uniform distribution. Consequently the joint state $C_A C_B$ of the measured two half-pairs together will be either 00 or 11.

2. If $A_1 = C_A$ then she sends $A_2$ to Bob over the classical channel acknowledging the fact that Bob's measurement result equals Alice's first bit $C_B = A_1$, i.e., the value of $A_1$ has been transmitted to Bob via the entangled pair. Furthermore, if Bob sent no bit to the channel then collision has been avoided and thus $A_2$ reaches Bob.
3. If $A_1 \neq C_A$ then she sends nothing to Bob. In this way Bob will be informed indirectly that $C_B \neq A_1$, or more precisely $C_B = \overline{A_1}$, therefore he has to invert $(0 \leftrightarrow 1)$ $C_B$ to get the proper value of $A_1$.

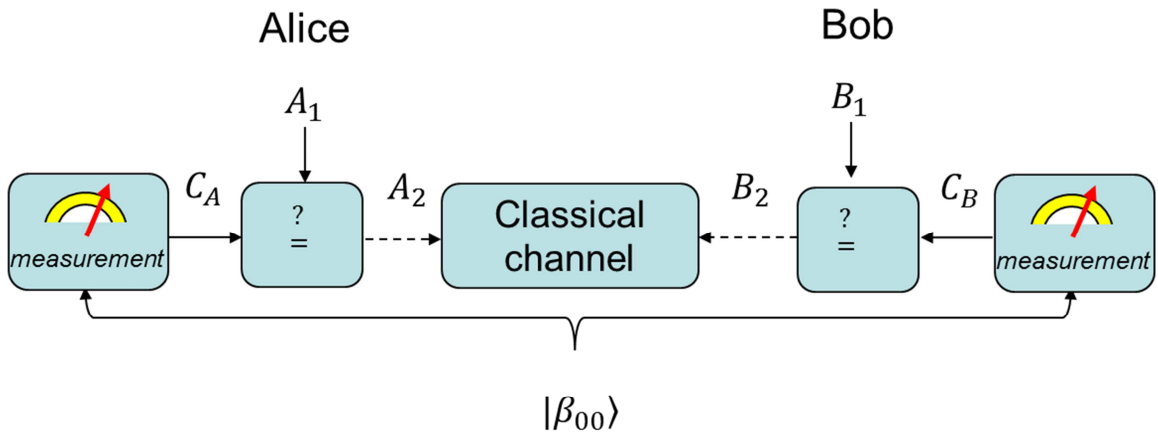

Figure 3.1 Hyperdense coding

Bob has to follow the same protocol steps at the same time. The possible values of the above discussed variables are summarized in Table 3.1. Each line represents a certain scenario indexed by *l* and sorted by possible values of $A_1$ and $B_1$.

If source bits of Alice and Bob are generated equiprobably and we know that $|\beta_{00}\rangle$ consists of two basis states with the same probability amplitudes then each of the eight scenario will occur with probability $p = 1/8$.

| *l* | $A_1$ | $B_1$ | $C_A C_B$ | Alice sends to the channel | Bob sends to the channel | channel state | successfully transmitted bits | number of successfully transmitted bits $K_l$ |
|---|---|---|---|---|---|---|---|---|
| 1 | 0 | 0 | 00 | $A_2$ | $B_2$ | Collision | $A_1, B_1$ | 2 |
| 2 | 0 | 0 | 11 | - | - | Unused | $A_1, B_1$ | 2 |
| 3 | 0 | 1 | 00 | $A_2$ | - | Transm. | $A_1, A_2, B_2$ | 3 |
| 4 | 0 | 1 | 11 | - | $B_2$ | Transm. | $A_1, B_1, B_2$ | 3 |
| 5 | 1 | 0 | 00 | - | $B_2$ | Transm. | $A_1, B_1, B_2$ | 3 |
| 6 | 1 | 0 | 11 | $A_2$ | - | Transm. | $A_1, A_2, B_2$ | 3 |
| 7 | 1 | 1 | 00 | - | - | Unused | $A_1, B_1$ | 2 |
| 8 | 1 | 1 | 11 | $A_2$ | $B_2$ | Collision | $A_1, B_1$ | 2 |

Table 3.1 Operation of hyperdense coding

Now, we can calculate the expected number of the successfully transmitted bits between the parties

$$E(K_l) = \sum_{l=1}^{L} p\, K_l = \frac{20}{8} = 2{,}5. \tag{2.7}$$

I one compares this 2,5 bits per timeslot with the performance of original slotted-Aloha system which can deliver for $M = 2$ users 0,5 bits in average according to (2.5) the improvement is considerable. When analysing the operation of the proposed protocol it is worth emphasizing the reason behind this increase. As we can observe in Table 3.1 Alice and Bob are playing with bits $A_2$ and $B_2$ a similar protocol to slotted-Aloha. These bits appear in the channel and collide or will be delivered successfully with the same probabilities as in case of slotted-Aloha. However, the same statistics comes not from the random behaviour of the parties but from the protocol itself. Therefore the bits $A_2$ and $B_2$ carry not more information in average than 0,5 bits in each timeslot. Where are the further 2 bits delivered per timeslot? The bits $A_1$ and $B_1$ are encoded into the presence of bits $A_2$ and $B_2$ in the channel by means of entanglement. Without entanglement

Now, we compare the new protocol to superdense coding which is able to deliver 2 classical bits in each timeslot [4]. On one hand hyperdense coding looks outperforming superdense coding if one considers 2,5 bits versus 2 bits while on the other hand because of the two-way nature of the protocol only 1,25 bits is transmitted in one direction between Alice and Bob which remains under the efficiency of superdense coding.

Furthermore, hyperdenese coding works only statistically while superdense coding provides 2 bits per timeslot deterministically. Finally, superdense coding requires joint measurement of the entangled qubits while in case of hyperdense coding Alice and Bob can measure their half-pair independently.

The proposed hyperdense coding protocol is fair in terms of Alice and Bob are being able to transmit information with the same probability, i.e. they can deliver statistically the same amount of information. This important property is ensured by entanglement which results in 00 and 11 with the same probability when measured.

4. Conclusions

In this paper we introduced a combined classical-quantum protocol for distributed access which delivers 2,5 classical bits in average in each timeslot using a previously shared entangled pair. Comparison of its efficiency were given to superdense coding and slotted-Aloha protocol.